\documentclass[12pt,preprint]{aastex}

\newcommand{\myemail}{quanz@mpia.mpg-hd.de}

\slugcomment{To appear in ApJ}

\shorttitle{FU Orionis - The MIDI/VLTI Perspective}
\shortauthors{Quanz et al.}

\begin{document}


\title{FU Orionis - The MIDI/VLTI Perspective\footnote{Based on observations made 
with ESO Telescopes at the Paranal Observatories under program ID 074.C-0209 
and 274.C-5032}}


\author{S. P. Quanz, Th. Henning, J. Bouwman, Th. Ratzka, Ch. Leinert}
\affil{Max Planck Institute for Astronomy, K\"onigstuhl 17, 69117 Heidelberg,
    Germany}
\email{\myemail}






\begin{abstract}
We present the first mid-infrared interferometric 
   measurements of FU Orionis. We clearly resolve structures that are best 
   explained with an optically thick accretion disk. A simple accretion disk
   model fits the observed SED and visibilities reasonably well and does not
   require the presence of any additional structure such as a dusty envelope. 
   The inclination and also the position angle of the disk can be 
   constrained from
   the multibaseline interferometric observations. Our disk model
   is in general agreement with most published near-infrared interferometric measurements.
   From the shape and 
   strength of the 8-13$\mu$m spectrum the dust composition 
   of the accretion disk is derived for the first time. We conclude that 
   most dust particles are amorphous and already much larger than those
   typically observed in the ISM. Although the 
   high accretion rate of the system provides both, high temperatures out to large radii and 
   an effective transport mechanism to distribute crystalline grains, we do not see any evidence for 
   crystalline silicates neither in the total spectrum nor in the 
   correlated flux spectra from the inner disk regions.
   Possible reasons for this non-detection are mentioned. All results
   are discussed in context with other high-spatial resolution 
   observations of FU Ori and other FU Ori objects. We also address the question
   whether FU Ori is in a younger evolutionary stage than a classical TTauri star.

\end{abstract}





\section{Introduction}
FU Orionis is the prototype of a small, but quite
remarkable class of low-mass Young Stellar Objects (YSOs) normally
referred to as FU Ori objects (FUORs). For the first members of
this class an outburst in optical light of up to 4-6 magnitudes
over short time scales, followed by a decrease in luminosity over
several years or decades, was observed \citep{herbig1977}. Other objects were
included in the class as they shared common specific spectroscopic
features, e.g., double-peaked line profiles, a spectral type
varying with wavelength and often CO bandhead absorption
features. As for one object (V1057 Cyg) the pre-outburst 
spectrum is known and as it resembles that of a classical TTauri 
star \citep{welin},
it is commonly assumed that FUORs should be low-mass YSOs.
Most observational data can be
explained by the presence of an accretion disk surrounding the
young stars \citep[][however, found that some spectral properties are well explained in the context of rapidly rotating late-type stars with strong stellar winds]{herbig}. 
A dramatic temporal increase in the accretion rate, where the disk outshines the star
by several orders of magnitude, can account for the observed outbursts
in luminosity. Several scenarios, possibly triggering such an
increased accretion rate, have thus far been proposed. They include
(a) interactions of binary or multiple systems where tidal forces
disturb the circumstellar disk \citep{bonnell}, (b) planet-disk
interactions, where thermal instabilities in the disk are caused
by the presence of a massive planet \citep{lodato}, or (c) thermal
instabilities in the disk alone \citep{bell}. For a detailed overview concerning the FU Ori
phenomenon we refer to \citet{hartmann}. Apart from revealing
the mechanism leading to the observed outbursts, it is also
important to investigate whether all TTauri stars undergo such epochs
of enhanced accretion or whether FUORs are a special class of
YSOs. Most observations of classical TTauri stars show that the
derived accretion rates of $10^{-10}-10^{-7}M_{\sun}/yr$ 
\citep[e.g.,][]{gullbring} are not sufficient to build up a low-mass star 
over time scales of a few Myr. Even if part of the matter is
supposed to be accreted in the very early phases of a YSO,
FUOR-phases might provide an elegant solution to this problem as
they would speed up the accretion process.

As so far only a dozen or so well established FUORs are known and as
the FUOR-phase has significant impact on the young star-disk system 
it is crucial to combine as much observational information as possible for these objects 
to eventually derive a coherent theoretical picture.
Based on near-infrared (NIR) and/or mid-infrared (MIR) interferometry as a technique to study the
inner few AU of the accretion disks new insights to
some of the best studied FUORs were provided recently. \citet{millan-gabet} found that 
accretion disks alone can not reproduce the SED and observed low K-band visibilities 
for V1057 Cyg, V1515 Cyg and Z CMa-SE simultaneously. 
They concluded that additional uncorrelated flux may arise
due to scattering by large dusty envelopes. \citet{abraham} presented the 
first VLTI/ MIDI observations of V1647 Ori whose eruptive behavior 
suggests that it is either an FUOR or an EX Lupi (EXor) type object.
In this case it was possible to fit both, 
the SED and the observed MIR visibility with a simple disk
model with moderate disk flaring. 
For FU Ori itself it was \citet{malbet} who analyzed a wealth of NIR interferometric data.
They showed that the NIR visibilities and the SED could be fitted with 
two models: One consisting simply of an optically thick and 
geometrically thin accretion disk and a second one consisting of an 
accretion disk and an embedded "hot spot". From their error statistics 
these authors concluded that the latter model was more likely. 

In summary it becomes clear that up to now no coherent picture can be 
derived from the interferometric 
observations of FUORs and the group of objects seems to be rather inhomogeneous.

 
In this paper we present the first MIR interferometric measurements of FU Orionis.
The data are thus complementary to the NIR observations of Malbet et al. (2005).
In section 2 we briefly describe the observations and the data 
reduction process. In section 3 we 
discuss the findings derived from the N-band acquisition images. The
8-13$\mu$m spectrum of FU Ori is analyzed in section 4. In sections 5 and 6 we discuss
the results from the interferometric measurements, i.e. the visibilities and the
correlated flux spectra, respectively. A simple analytical disk model and its fit to the 
SED and the visibilities is presented in section 7. Finally, we summarize our conclusions and 
mention some future prospects in section 8.

\section{Observations and Data Reduction}


The observations were carried out between 
October 31$^{st}$ and November 4$^{th}$ 2004 with the Mid-Infrared Interferometric 
Instrument (MIDI) at ESO's Very Large Telescope Interferometer (VLTI)
on Paranal/Chile. Together with the Keck Interferometer Nuller that recently started to produce first scientific results
\citep{mennesson}, MIDI is currently worldwide the only instrument 
able to conduct spectrally resolved interferometric observations in the mid-infrared
\citep{leinert}. For the observations MIDI was used in high-sens mode using the NaCl prism  
as dispersive element yielding a spectral resolution of 30. The maximum projected baseline
was  86.25m and the minimum projected baseline 44.56m leading to an angular resolution at 10$\mu$m of 0.$''$029 and 0.$''$056, respectively. For an assumed distance of 450 pc these values correspond to 13.1 AU and 25.2 AU.
A journal of the observations including
projected baselines, position angles and calibrator stars is given in 
Table~\ref{journal}. For completeness we also mention the
observations from December 2004 although they are disregarded in the following sections.
These observations had almost exactly the same baseline and position angle as the
October observations but showed in general a lower level of total and 
correlated flux due to rather poor observing conditions in terms of seeing and transparency. 
The calibrator stars for the data reduction were chosen by analyzing all calibrator stars 
observed over the whole night. We selected those showing a good
agreement in their transfer functions, i.e. their instrumental visibilities after the
assumed sizes were taken into account.  

The data reduction was carried out with the 
software package MIA+EWS-1.3. This software consists of 
two independent reduction programs (MIA and EWS) which were both
applied for comparison. The software, further information
and manuals can be downloaded from the 
Internet\footnote{http://www.strw.leidenuniv.nl/$\sim$nevec/MIDI/index.html}.  
A general description of the basic data reduction steps
is also given in \citet{leinert} and we refer the reader to this paper. 
As the results derived with MIA and EWS agreed quite well (overall differences in the
correlated flux $\leq 8\%$), we decided to show only
plots resulting from the MIA package. 



\section{The MIDI Acquisition Image}


FU Ori was found to be a binary system by
\citet{wang}.  \citet{reipurth} confirmed the detected companion and
concluded that it was a young star of spectral type K showing clear 
NIR excess. In the same paper it was also stated that it
was very unlikely that the binary component (FU Ori S) triggered
the outburst of FU Ori observed in the late 1930s. 

\begin{figure}
\centering
    \epsscale{0.5}
   \plotone{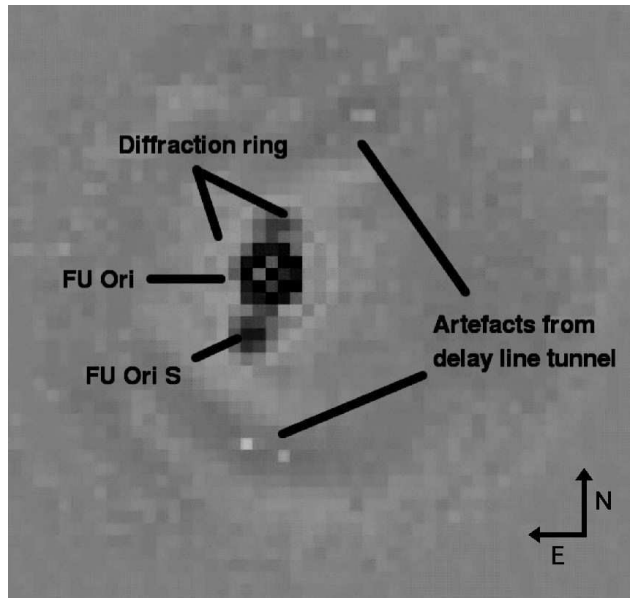}
   \caption{FU Ori and FU Ori S as seen in the MIDI acquisition images at 8.7$\mu$m. The image was
   taken with UT3 the 31st of October 2004 using the AO-system MACAO (north is up, east to the left).}
              \label{Image}%
\end{figure}

Knowing about the existence of the fainter companion the integration time
of some MIDI acquisition images was increased in order to derive
N-band photometry for both components. In three images FU Ori S was 
clearly visible (Figure~\ref{Image}) and aperture
photometry could be applied to the observations. The results are summarized in Table~\ref{fluxes}.
Interestingly, FU Ori S shows relatively a higher N-band excess
than FU Ori itself. This can be explained in terms of differences in the geometry of an assumed circumstellar 
disks (e.g., larger flaring angle or smaller disk inclination).
We furthermore derive a separation between the components of 0.$''$484$\pm$0.$''$01 and 
a position angle of 162.5$^\circ\pm$4.1$^\circ$ (measured from north eastwards)
which is in good agreement with the results from \citet{wang} and \citet{reipurth}. Our errors are standard deviations based on Gaussian fitting of 3 independent images. For an assumed distance of 450 pc the
separation corresponds to 217.8$\pm$4.5 AU.



\section{The Total Uncorrelated MIR Spectrum}
\subsection{Comparison to \emph{Spitzer} observations}
\begin{figure}
\centering
\vspace{1.5cm}
    \epsscale{0.6}
    \plotone{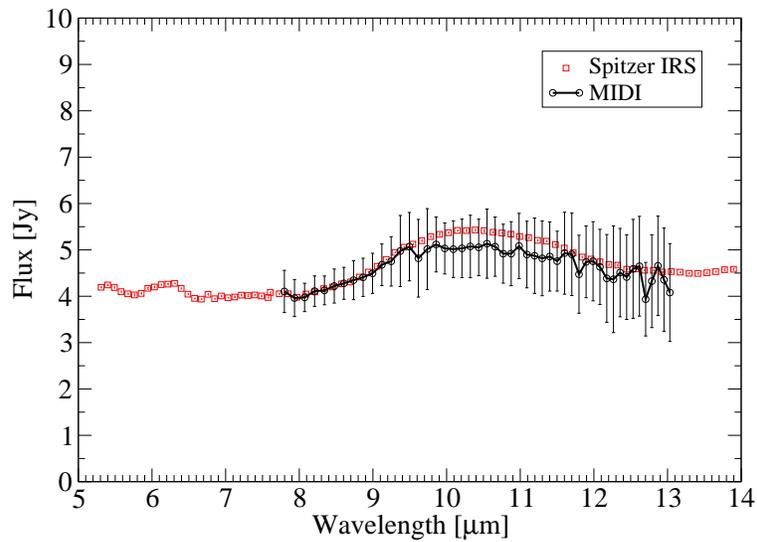}
   \caption{MIR spectrum of FU Ori from 5.3-14$\mu$m. Data from the MIDI observations 
   (dots with solid line) range from 8-13$\mu$m. They agree within 
   the error bars with  
   \emph{Spitzer IRS} data (squares). The errors for Spitzer are smaller 
   than the symbols and are not shown. The larger 
   MIDI errors arise from averaging six independently calibrated measurements (3 nights
   and 2 telescopes each night). The observed silicate feature is rather flat and broad.}
 \label{spectrum_short}%
\end{figure}

Figure~\ref{spectrum_short} shows the MIR spectrum of FU Ori. The spectrum obtained with 
MIDI is the average over the 
first three observing nights using both Unit Telescopes (UTs). For comparison 
\emph{Spitzer IRS} data are plotted 
additionally. Those data were available from the \emph{Spitzer} Data Archive. 
For the data reduction 
we used the \emph{droopres} intermediate data product processed through the SSC pipeline S12.0.2.
The SMART reduction package developed by the IRS Instrument Team at Cornell
\citep{higdon}
was used to extract the spectrum. Within the error bars the spectra agree quite well. 
A broad and weak 
silicate emission feature is present between 8-13$\mu$m. However, the averaged
MIDI spectrum appears to be a little flatter than the more roundish \emph{Spitzer} spectrum. 

\subsection{Possible dust composition}
In contrast to other YSOs the observed silicate emission feature of FU Ori is 
rather weak \citep[see also][]{hanner}. 
Such a broad and flat MIR dust spectrum can be explained by grains that already
underwent some coagulation process \citep{bouwman}. To test 
this hypothesis we derived a possible
dust composition by fitting a dust model to the normalized and continuum subtracted 
\emph{Spitzer} spectrum from 6-13$\mu$m. 
This analysis method is similar to approaches that were successfully applied in previous 
studies for determining the dust composition of observed dust emission features of other YSOs \citep[e.g.,][]{bouwman}.
In the model we assume that the observed emission can be computed from the sum of the
emission of individual dust species. The species we used are summarized in Table~\ref{dust_species} together
with references for their optical properties. We furthermore took into account different grain sizes 
as for each of the dust species the opacities were calculated for 0.1, 1.5, and 6.0$\mu$m sized particles.
To estimate the possible contribution of Polycyclic Aromatic Hydrocarbons (PAHs) to the emission 
we used a template spectrum based on profiles that were derived from observations \citep{peeters,diedenhoven}.
Finally, in order to reproduce the observed spectrum we fitted the following emission model to the
\emph{Spitzer} data using a least square minimization
\begin{equation} 
F_{\nu}-B_{\nu}(T_\mathrm{cont})C_0=B_{\nu}(T_\mathrm{dust})\Bigg(\sum^{3}_{i=1}\sum^{5}_{j=1}C_{i,j}\kappa_{\nu}^{i,j}\Bigg)
+C_\mathrm{PAH}I_{\nu}^\mathrm{PAH}\quad .
\end{equation}
Here, $F_{\nu}$ is the observed flux, $B_{\nu}(T_{cont})$ denotes the Planck function at the temperature of the underlying continuum, 
$B_{\nu}(T_{dust})$ the Planck function at the temperature of the silicate grains, $\kappa_{\nu}^{i,j}$ is the mass absorption
coefficient for the silicate species $j$ with the grain size $i$, $I_{\nu}^{PAH}$ is the template emission spectrum for the PAHs
, and $C_0$, $C_{i,j}$, and $C_{PAH}$ are the weighting factors for the continuum, the silicate emission and the PAH 
contribution, respectively. A more detailed description of the dust model 
and plots showing the different dust opacities is given in \citet{bouwman2006}. 

\begin{figure}
\centering
    \epsscale{0.6}
   \plotone{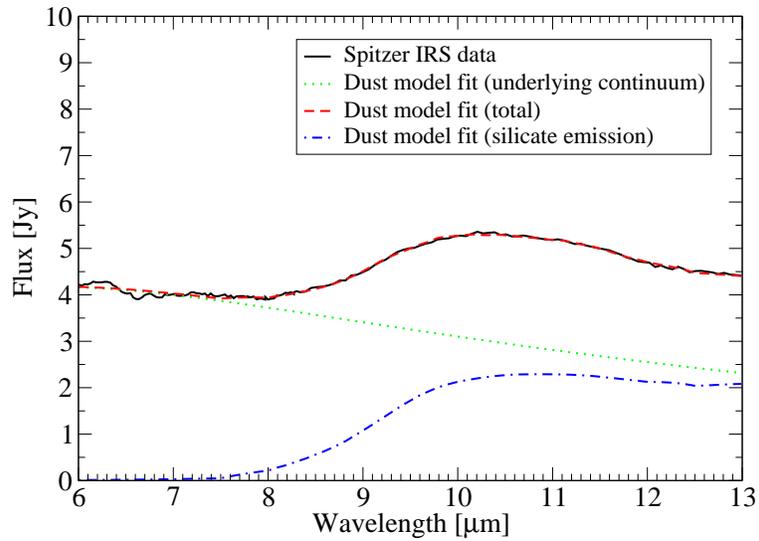}
   \caption{MIR spectrum of FU Ori and the resulting dust model fit. The \emph{Spitzer} 
   spectrum (black solid line) can be fitted quite well with the dust composition 
   given in Table~\ref{dust_compo}. The dotted line shows the contribution of the
   underlying continuum, the dash-dotted line shows the silicate emission and
   the grey dashed line shows the sum of both, i.e. the total flux, which is hardly
   distinguishable from the observed spectrum.}
 \label{dust_spectra}%
\end{figure}

The model results are shown in Figure~\ref{dust_spectra} and Table~\ref{dust_compo} summarizes the
corresponding dust composition.
As for the underlying 
continuum we could fit a temperature of 880 K, we believe that  
this optically thick emission most likely arises from the inner parts of the disk close to the star.  
The outer and thus cooler regions of the disk would then be 
responsible for the weak silicate emission 
for which we fitted a temperature of 230 K to the optically thin emission layer. 
The results strongly support the 
idea that most particles are amorphous and much larger in size than those in the ISM. PAHs, silica particles and 
crystalline silicates do not seem to be present at all. Although the derived fractions of different dust species do depend on the
model applied to the data (e.g., in terms of the precise grain sizes considered and also in terms of particle structures) the main conclusions, that we
do not find evidence for crystalline silicates and that the grains are considerably larger than those found in the ISM, remain unaltered. For a more detailed discussion about the influence of different model parameters we refer to \citet{voshchinnikov} and \citet{boekel}.


As the disk is much brighter in the optical and NIR than the central star (section 7),
stellar radiation is probably not able to produce the observed weak emission feature
in an optically thin disk surface layer. However, it was shown that the silicate 
feature of FU Ori can be reproduced taking into account self-irradiation of the disk \citep{lachaume}.
In this case the hot inner parts of the accretion disk serve as flux 
source and illuminate the disk surface layer under a certain angle.

\section{The Visibilities}
\begin{figure}
 \vspace{1.5cm}
\centering
    \epsscale{0.6}
   \plotone{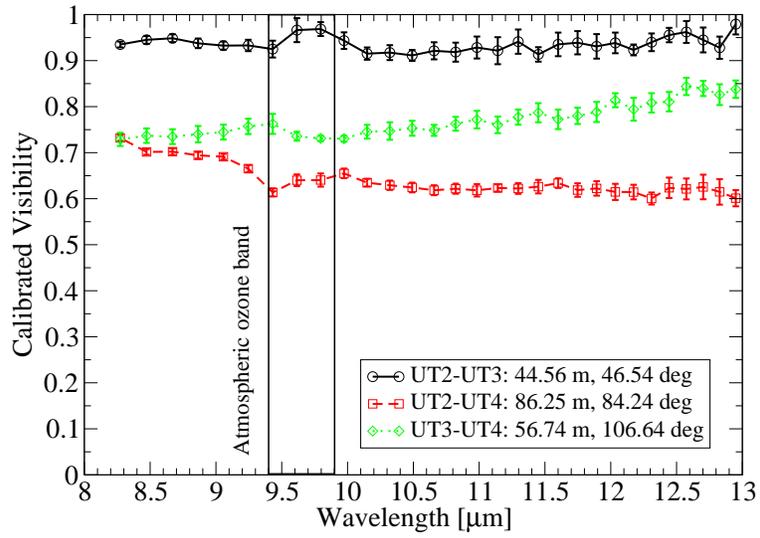}
   \caption{Calibrated Visibilities of FU Ori for three different baselines.}
 \label{visibilities}%
\end{figure}

In Figure~\ref{visibilities} and Table~\ref{visibilities_table} the visibilities measured for FU Ori at
three different baselines are shown. 
The errors are computed as standard deviations resulting from the use of different calibrators for
each night (see Table~\ref{journal}). 
In most of the following plots the position of the atmospheric ozone band is indicated 
as this part of the spectrum suffered sometimes from imperfect 
corrections during the data reduction process.  Here, small ``dips'' and 
``bumps'' in otherwise flat spectra are remnants from the data 
reduction. 

\subsection{Qualitative assessment}
The calibrated visibility of the shortest
baseline (UT2-UT3) remains almost constant at $\approx$0.95 over
the whole wavelength range and the source is only 
marginally resolved. As expected, for the longest
baseline (UT2-UT4) the lowest visibility is observed ranging
from $\approx$0.73 at 8.3$\mu$m to $\approx$0.6 at 13$\mu$m.
For  the intermediate baseline (UT3-UT4) the visibility shows a 
slight increase from $\approx$0.73 at
8.3$\mu$m to $\approx$0.85 at 13$\mu$m. 
The fact that the object is 
clearly resolved with two baselines in the MIR supports the argument 
of \citet{malbet} that 
we are observing an extended circumstellar structure, i.e. a disk, and 
not only the stellar 
photosphere \citep{herbig}. 

For the two baselines UT2-UT3 and UT2-UT4 the observations are consistent
with expectations from thermal disk emission as despite the decreasing resolution
for longer wavelengths the visibilities indicate larger sizes for the emitting regions (see also section 5.3.). 
The observed increase in visibility for the
UT3-UT4 baseline implies that for this
baseline, however, the object appears smaller at 13$\mu$m than at 8$\mu$m.
This seems difficult to imagine in the context of a
circumstellar disk as emitting source and certainly a second observation 
for this baseline configuration seems eligible. Theoretically it is possible that the 
photometric measurements carried out directly after the interferometric
observations are corrupted due to technical problems or different weather conditions. 
This in turn might then lead to a change in the visibility function. However,
the calibrators observed at this night did not show any sign of poor
photometric measurements and their transfer function was very stable over the whole night. 
In addition, also other YSOs with circumstellar disks showed an
increasing visibility for certain baselines when observed with MIDI. 
Thus, since we do not find any evidence for excluding this dataset 
due to bad quality we decided to keep it in our analyses.

\subsection{Comparison to MIR visibilities of HAeBe stars}
As thus far no MIDI visibilities for TTauri stars have been published we are limited 
to a comparison between FU Ori and 
circumstellar disks around Herbig Ae/Be stars (HAeBes). 
It shows that these stars
as well as models applied to them normally show a prominent drop
in the visibility between 8$\mu$m and 10$\mu$m from where the curve remains almost
constant \citep{leinert}. 
Qualitatively, this drop results from the
intensity distribution of the passive disks around the HAeBes: At the short wavelength end
the hot inner rim of the disks provides an overproportional contribution to the
flux and is at the same time confined to a small spatial region leading to a high visibility.
Most of the rest of the MIR emission originates from a 
large area of the hot, illuminated 
surface layer of the flared circumstellar disk.
The visibilities for FU Ori show hardly any wavelength dependence and 
are very flat from 8-13$\mu$m regardless of baseline. Thus, its seems as if 
the flux distribution of FU Ori is smoother and the visibilities show 
no sign of a significant contribution from a hot inner rim.  
However, as FU Ori is surrounded by a heavily active accretion disk where 
the disk alone produces the majority of the observed flux at almost all wavelengths 
(see section 7) differences in the
intensity distribution and thus in the visibility can be expected.

Apart from differences in the shape of the visibility curves most HAeBes are much better
resolved, i.e. show lower visibilities than FU Ori \citep{leinert}. This, however, can 
at least partly be explained by the distance to these objects which is in general less 
than 200 pc. FU Ori on the other hand has an assumed distance of 450 pc, and 
if it was closer to the Earth we would observe lower 
visibilities also for this object.

\subsection{Geometry of the emitting regions}
\begin{figure}
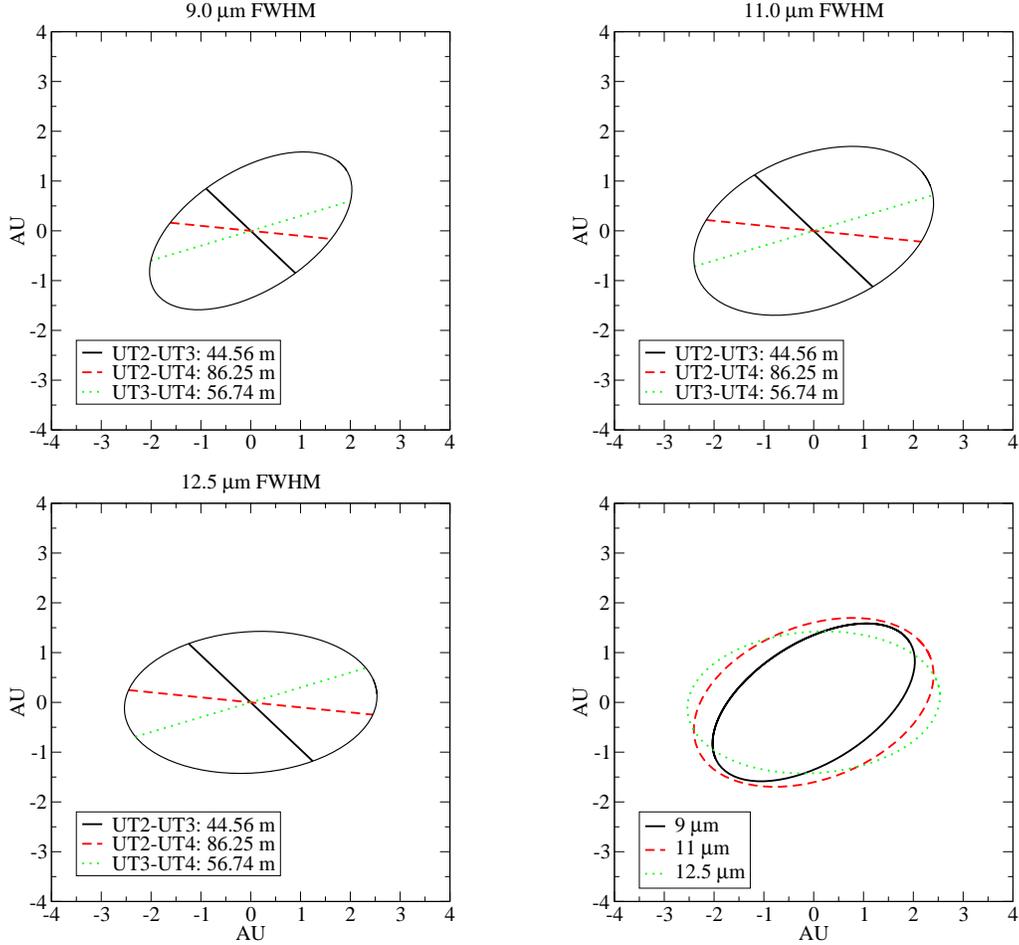

\centering
\epsscale{0.8}
    \plottwo{f5a_color.eps}{f5b_color.eps}
    \plottwo{f5c_color.eps}{f5d_color.eps}
   \caption{Sizes and orientation of the emitting regions on the sky for different 
   wavelengths (9.0$\mu$m (upper left), 11.0$\mu$m (upper right) and 12.5$\mu$m (lower left)).
   The sizes 
   correspond to FWHM of assumed Gaussian brightness distributions and are 
   given in Table~\ref{table_sizes}. The resulting best fit ellipses for each 
   wavelength are overplotted and combined in a fourth plot (lower right). The center
   of the plots corresponds to the position of FU Ori and north is up and east to the left.}
 \label{figure_sizes}
\end{figure}

To derive a simple model for the
geometry of the emitting regions we assume a simple Gaussian brightness distribution for 
each baseline. This is a reasonable first approximation for objects showing high visibilities. 
The FWHM, and hence the physical size, of this Gaussian in arcsec can be computed by 
\begin{equation}\label{visi_eq}
\Theta = \sqrt{\frac{ln(V(f))}{-3.56\cdot f^2}}
\end{equation}
where $V(f)$ is the measured visibility for a certain spatial frequency $f=\frac{B}{\lambda}$ (in arcsec$^{-1}$) 
derived from the projected baseline $B$ and the wavelength $\lambda$.
Equation~\ref{visi_eq} results from a simple Fourier transformation of the assumed brightness
distribution.
We computed the FWHM for all three baselines at three different 
wavelengths (9.0, 11.0, and 12.5$\mu$m). For this we averaged for each baseline 5 visibility points from Table~\ref{visibilities_table} centered on the specified wavelengths. 
Table~\ref{table_sizes} gives the resulting sizes of the emitting regions in AU for
an assumed distance to FU Ori of 450 pc.
The results are also visualized in Figure~\ref{figure_sizes} where the
FWHM are shown in their orientation on the sky. As expected for thermal disk emission the FWHM
increases with wavelength for a given baseline. The only exception is
the 12.5$\mu$m size for the UT3-UT4 baseline which is 
surprisingly a little smaller than that seen at 11.0$\mu$m. 


In addition to size estimations for each individual baseline our measurements based on  
different position angles allow to constrain the geometry of the disk. 
We fitted the derived FWHM
with an ellipse for each of the considered wavelengths in order to derive a simple model for 
the spatial orientation of an assumed disk-like structure. The resulting best fit ellipses are 
overplotted in Figure~\ref{figure_sizes} and their parameters are summarized in 
the lower half of Table~\ref{table_sizes}. 
As one would expect from a disk-like structure
showing a decrease in temperature with radius the semimajor axes of the ellipses increase
with wavelength. However, due to the increasing visibility of the UT3-UT4 baseline the semiminor 
axis and also the total area of the ellipse are slightly smaller at 12.5$\mu$m than at 11.0$\mu$m.

An inclination angle for 
an assumed circular disk can be derived by 
computing the $arccos$ of the ratio of the semiminor and semimajor axis. 
The resulting angles agree very well 
(55.4$^\circ\pm$2.4$^\circ$) and are also in good agreement
to what was found for FU Ori disk models based on NIR observations \citep{malbet}. 
By combining the derived inclination angle with rotational velocity measurements
in the optical spectral region it is possible to estimate the central mass. Based on 
\citet{kenyon1988} we derive for FU Ori $M/M_{\sun}\approx\,$0.36 which is reasonably consistent
with typical values for low-mass pre-main sequence stars.

In addition to the inclination the fitted ellipses provide also information 
on the position angle of the disk.  
The position angles (measured from north eastwards) we find for the semimajor axes at different wavelengths 
show a larger scatter (109.1$^\circ\pm$11.6$^\circ$) than the inclinations  
and differ significantly from the NIR findings of \citet{malbet} 
(see also section 7). It seems as if the position angle becomes smaller for longer wavelengths,
although one has to keep in mind that, again, at least for the UT3-UT4 baseline the increasing visibility 
is the main reason for the observed rotation. 
At this point we leave it to future investigations to derive a 
3-D disk model based on a more complex dust distribution possibly
able to confirm the apparent changes in the position angle for different wavelength regimes. 

\section{The Correlated MIR Spectra}
\subsection{The origin of the correlated flux}
\begin{figure}
\centering
    \epsscale{0.6}
   \plotone{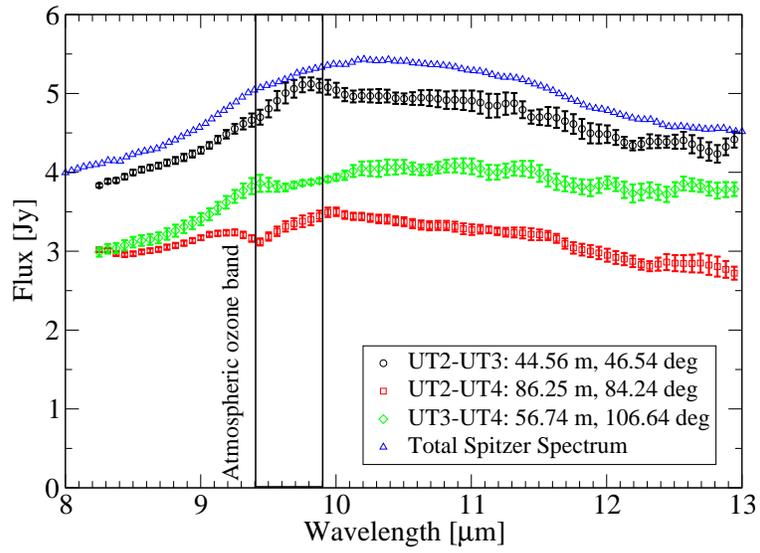}
   \caption{Observed correlated flux for the three different baselines and 
   position angles.}
 \label{correlflux}%
\end{figure}

The correlated flux is directly linked to the total flux (i.e. the flux from 
a single UT telescope) via the visibility:
\begin{equation}\label{correlflux_eq}
F_{corr}(\lambda)=V(\lambda)\cdot F_{total}(\lambda)
\end{equation}
Figure~\ref{correlflux} depicts the results for the correlated flux for 
the three different baselines.
The total spectrum is plotted for comparison. 

Taking into account the spatial resolution of the three baselines it is
possible to estimate from where the correlated 8-13$\mu$m flux originates.
Our UT2-UT3 baseline has a spatial resolution of $\approx$25 AU at the
distance of FU Ori, and if we assume an average visibility of 0.95
we hence know that 95\% of the 8$\mu$m-13$\mu$m flux must come from 
within this 25 AU. Similarly, we find that for the UT2-UT4 baseline 65\% of the flux is emitted from 
the inner 13 AU. For the UT3-UT4 baseline 75\% of the 
8-13$\mu$m flux comes from the inner 20 AU. 

\subsection{Lacking crystalline silicates?}
The spatially resolved MIR spectra of the MIDI instrument make it
possible to study the radial dependence of the dust composition in the protoplanetary 
disk \citep{boekel}. In section 4 we showed that the spatially unresolved (total) spectrum of FU Ori 
can be fitted with mainly large amorphous dust grains. The shape of the
spatially resolved (correlated) flux spectra contains information about
the dust composition of the inner parts of the disk.

\begin{figure}
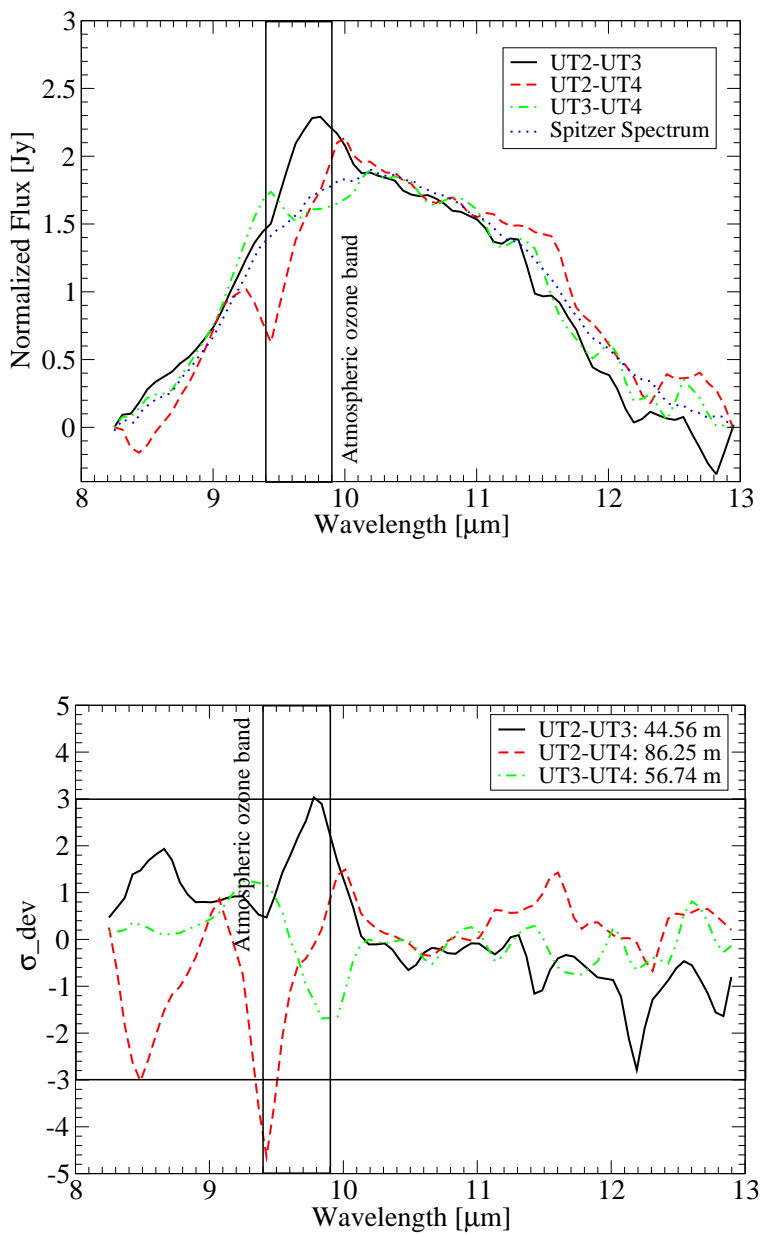

\centering
    \epsscale{0.6}
    \plotone{f7a_color.eps}
 \vspace{2cm}
     \epsscale{0.6}
    \plotone{f7b_color.eps}
   \caption{\emph{Upper Panel:} Continuum subtracted and normalized correlated spectra for the 
   three different baselines and 
   the normalized total spectrum. For clarity the errorbars have not been overplotted. They equal those of
    Figure~\ref{correlflux}. \emph{Lower Panel:} Variations of the  normalized correlated spectra with respect to 
   the normalized total \emph{Spitzer} spectrum in units of error. It shows that most variations 
   are clearly below the 3$\sigma$ threshold, indicating that the shape of the correlated spectra equals that 
   of the total spectrum.}
 \label{shape_correl}%
\end{figure}

To compare the shape of the correlated spectra to the total spectrum we first subtracted the
continuum for which we fitted a straight line between the flux at 8.25 and 12.95$\mu$m.
Then we normalized all continuum subtracted correlated flux spectra and the total \emph{Spitzer} spectrum by
\begin{equation}
F_{norm}=\frac{F_{sub}}{F_{mean, sub}}\quad 
\end{equation}

where $F_{sub}$ and $F_{mean, sub}$ denote the continuum subtracted spectra and their mean value, 
respectively. By applying this formula all correlated spectra conserve 
their shape and can be compared to the total spectrum (Figure~\ref{shape_correl}). 
To identify those parts where the spectra differ significantly we furthermore computed 
the deviation from the total \emph{Spitzer} spectrum
for each wavelength in units of the data error:
\begin{equation}
\sigma_{dev}=\frac{F_{norm, corr}-F_{norm, total}}{\sigma_{corr}}
\end{equation}

$F_{norm, corr}$ and $F_{norm, total}$  are the normalized correlated spectra and the normalized total spectrum
and $\sigma_{corr}$ denotes the standard deviation of the correlated 
spectra (MIDI errors) at a given wavelength.
We applied the same normalization factors to the errors as we did to the spectra.
Figure~\ref{shape_correl} shows that 
no clear deviations from the total unresolved spectrum are observed
for any of the correlated spectra. This means that with the given spatial resolution 
and sensitivity no significant chemical 
difference in the dust composition is observed and the spectra of the
inner parts of the disk look very similar to the total unresolved disk spectrum.  

This finding is somewhat surprising. 
As seen in section 4 the silicate dust grains apparently already underwent noticeable
coagulation. Thus, the disk does not consist of purely ISM dust anymore and must already have a
significant age. In addition it is known that apart from 
grain growth also thermal annealing takes place within protoplanetary
disks and transform the amorphous silicates into crystalline silicates \citep[see,][]{henning2006,ancker}.
However, neither the total 
spectrum nor the correlated spectra indicate the presence of crystalline silicates.
This is
even more surprising considering the high temperatures in the disk interior
due to viscous heating of this actively accreting star (section 7). Annealing processes
transforming ISM amorphous dust into crystalline dust set in in disk regions where $T\ge 800K$
\citep[e.g.,][]{gail1}. In addition to that, high accretion rates directly
relate to a faster vertical and horizontal mixing within the protoplanetary disk \citep{gail2}. This
effect also eases the detection of crystalline silicates within the accretion
disk surrounding FU Ori. But why don't we find any evidence for their existence?

\begin{figure}
\centering
    \epsscale{0.6}
   \plotone{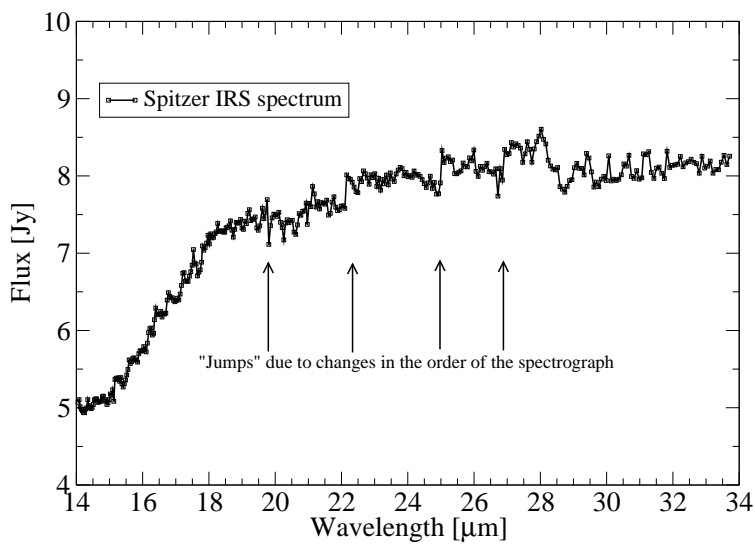}
   \caption{\emph{Spitzer} MIR spectrum from 14-34$\mu$m. The spectrum was smoothed by a factor of three.
   The typical errors are smaller than the size of the symbols. Similar to 
   the shorter wavelength spectrum in Figure~\ref{spectrum_short} 
   this spectrum appears rather smooth without any striking features. The arrows 
   indicate where the order of the spectrograph is changing leading to variations in the
   detected flux (approximately at 19.6, 22.1, 25.0, and 26.7$\mu$m).  
   Although the weak peak around 27.7-27.8$\mu$m can theoretically be 
   attributed to the presence of forsterite particles their detection seems 
   questionable as the normally stronger emission bands at 23.9 and 33.8$\mu$m 
   is clearly not seen.
   }
 \label{spectrum_long}%
\end{figure}

A possible explanation might be that they simply 
cannot be detected as the related feature is rather weak and simply dominated 
by the underlying continuum. This hypothesis can be checked by analyzing the 
longer wavelength regime of the \emph{Spitzer} spectrum as shown in Figure~\ref{spectrum_long}.
As at longer wavelengths the temperature of the continuum is decreasing the
contrast between the continuum and possible emission features due to the presence of
crystalline silicate particles increases. Hence, spectral signatures of these particles should
be more easily detectable. Figure~\ref{spectrum_long}, however, shows that also between 
14 and 34$\mu$m the spectrum of FU Ori does not show any clear signs of crystalline silicates. 
Normally crystalline forsterite particles have emission bands at 16.4, 23.9, 27.7 and 33.8$\mu$m and enstatite grains show features
at 18.8, 21.5, and 24.5$\mu$m \citep{molster}. If one argued that possibly around 27.7-27.8$\mu$m 
a weak forsterite feature was present one should keep in mind that the features at 23.9 and 33.8$\mu$m are normally
stronger \citep{min} but are not seen at all. Thus, also the longer wavelength MIR spectrum of FU Ori shows that crystalline silicates are not present. This shows that apparently the non-detection is at least not 
only due to contrast problems between the emission features and the underlying continuum.

What else can prevent us from detecting crystalline silicates? In case FU Ori underwent its 
first (of possible multiple) outburst(s) and phase(s) of enhanced
accretion, it is possible that not enough time has yet elapsed 
for the crystalline silicate particles to reach the disk surface
from where they can be detected by means of MIR spectroscopy. 
This hypothesis is, however, questionable as the large sizes of the amorphous silicates indicate 
that enough time, at least for dust coagulation, has elapsed.
Recently, \citet{dullemond} found that assuming the crystallization by thermal annealing
happens at the very early phases of disk evolution, the level of crystallinity is linked
to the properties of the molecular cloud core from which the disk formed. 
It was found that rapidly rotating cores produce rather massive disks with a high accretion rate 
but a low level of crystallinity. As we observe these properties for FU Ori the proposed model, although yet very simple and with limited predictive power, might provide an interesting way to explain at least partly the lack of crystalline silicates in 
the FU Ori accretion disk.  
Finally, another possibility might be that the vertical and radial 
mixing within the disk does not work as efficient as expected. In consequence not enough 
crystalline dust grains reach the disk surface at radii where also the contrast between the
emission feature and the dust continuum allows 
their detection.

\subsection{Comparison to TTauri stars and HAeBe Stars}
The results from the correlated flux allow a comparison to disks
around other YSOs that have also been 
observed with VLTI/MIDI. Not only seem crystalline silicates be commonly present 
in protoplanetary disks, but 
for three intermediate-mass HAeBes stars \citet{boekel} found
that the degree of crystallization within the disk depends on the distance from the 
central star. The inner parts of the disks are more crystalline than the outer disk regions.
This can naturally be explained by assuming that in the inner disk regions where the
disk has higher temperatures dust processing and annealing occurs and that over time
crystalline particles are transported outwards. 
Not only for intermediate mass stars but also for the 
TTauri star RY Tau the same radial dependence of the crystallinity 
was observed \citep{schegerer}. Thus, crystalline silicates should in general 
be more easily detectable in the correlated spectra of protoplanetary disks. 
For reasons mentioned in the previous section it seems thus surprising that the disk 
around FU Orionis does not seem to contain any detectable amount of crystalline dust.

As described in the introduction FUORs are believed to be low-mass or TTauri-like 
stars in a state of enhanced accretion. From sub-millimeter and millimeter observations \citet{weintraub} concluded that
FUORs are possibly in an evolutionary youger age than classical TTauri stars. 
This younger age is also supported by theoretical models where the protostellar disk
is continuously replenished by a dusty envelope. This material infall on the disk 
can lead to thermal instabilities in the disk and thus provides a possible mechanism 
for explaining the observed outbursts \cite{bell}. 
The puzzling thing is now that apparently in the disk surrounding FU Ori 
the amorphous grains have already grown considerably. Hence, the
disk can not be in its native dust composition and might already have a significant age. 
Furthermore, from the \emph{Spitzer} FEPS (\emph{Formation and Evolution of Planetary Systems}) legacy program
it is known that apparently the amount of crystalline silicates in TTauri disks is not 
correlated with the average size of amorphous dust particles \citep{bouwman2006}. 
Hence, crystalline dust particles are
created in the \emph{early} phases of disk evolution where the accretion rate and 
the disk temperature are thought to be higher. If thus the FUOR-phase is indeed common
to most/all TTauri stars then the non-detection of crystalline particles cannot be explained
in the current picture of protoplanetary disk evolution.

\section{A Simple Disk Model}
\subsection{The SED}
The present paper is more intended to show the observational results
rather than to present a sophisticated disk model possibly able to 
match the observed SED and visibilities with highest accuracy. Thus, instead of applying 
a complex radiative transfer model we decided to use in a first step 
a simple analytical disk model. 
In this model the effective temperature and also
the surface density are prescribed by the following broken power law distributions

\begin{equation}
T(r)=T_{1AU}\Big(\frac{r}{1AU}\Big)^{-q}
\end{equation}
\begin{equation}
\Sigma(r)=\Sigma_0\Big(\frac{r}{r_0}\Big)^{-p}
\end{equation}

\begin{figure}
\centering
    \epsscale{0.6}
   \plotone{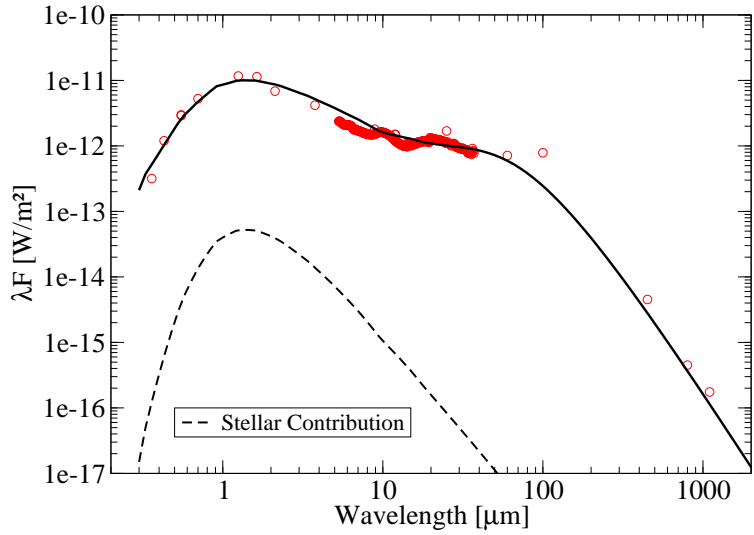}
   \caption{SED fit of a geometrically thin, optically thick accretion disk model. 
   The disk parameters are shown in Table~\ref{model_parameter}. The contribution of the central star
   (in this case we assume $T_{eff}=3700~K$, $R_{star}=2.5 R_{\sun}$) can be neglected. 
   Although simple, the disk model agrees 
   reasonably well with the observed fluxes for most wavelength regimes. The data points were taken from
   \citet{clarke} (U,B,V,R), \citet{przygodda} (8.9$\mu$m, 11.9$\mu$m), \citet{reipurth} 
   (J, H, K, L), \citet{weintraub} (450$\mu$m, 850$\mu$m,
   1.3mm), \emph{Spitzer IRS} data archive (MIR Spectrum) and \emph{IRAS} data archive (12~$\mu$m, 
   25~$\mu$m, 60~$\mu$m, 100~$\mu$m).}
 \label{sed}%
\end{figure}

For the temperature law the temperature at 1 AU and the
exponent $q$ are free parameters. For the
surface density the total disk mass $M_D$ and the exponent $p$ are free parameters. 
As the integrated surface density must equal the total mass of the 
disk the radial dependence of the surface 
density $\Sigma(r)$ can be computed. 
The shape of the total SED (and finally also the observed visibilities) require a double power-law to fit
the data. At a radius of 3 AU the temperature and surface density distribution
change. However, we made sure that for both parameters the transition 
was smooth and continuous. 
Figure~\ref{sed} shows the SED and the resulting
fit from the simple disk model. The model 
parameters are summarized in Table~\ref{model_parameter}. 
Based on the value for the visual extinction $A_{V}$ we 
interpolated the extinction for the other wavelengths following the 
findings of \citet{mathis}.
The disk inclination $i$ was derived from
the ellipses fitted to the FWHM of assumed Gaussian brightness distributions (see 
section 5.3.). For the position angle $\theta$ we used two values: One was
also derived in section 5.3., and for the second one we fitted the disk model to 
the observed visibilities with $\theta$ being the only free parameter (see also section 7.2.).

It is clear that this simple approach is not
able to reproduce all features of the SED (e.g., the weak
silicate emission feature). For the moment, however, we 
restrict ourselves to this simple approach and, as mentioned above, a detailed modeling is left for
further investigations. It is striking that rather high disk 
temperatures are required to fit the SED. In the inner 
0.35 AU no dust can withstand the high temperatures  $>$1500 K even 
on the disk surface and only the gaseous disk 
component can survive. It is interesting to note that the
power law index of $q1=0.75$ of the inner disk region equals exactly that what was derived in
analytical disk models for geometrically flat, optically thick \emph{passive} disks \citep[e.g.,][]{adams} 
as well as for
flat, steady state, optically thick \emph{accretion} disks \citep[e.g.,][]{pringle}. Thus, 
from the temperature distribution alone one can not distinguish between the two different cases 
\citep[see also][]{kenyon}. However, in the context of FU Ori the passive disk model where the disk 
is merely re-processing radiation from the central star is unlikely and only an accretion disk model
matches most observations \citep{hartmann}.
A nice example is the recently detected magnetic field in the
innermost regions of the disk surrounding FU Ori which can only be explained in the picture of a 
heavily accreting disk \citep{donati}. 
As the disk temperature is directly related to the mass accretion rate $\dot M$
in classical accretion disk models it is possible to derive the $\dot M$ required 
to produce the observed high temperatures. \citet{bell2} 
analyzed the effective disk temperature and also the
midplane temperature of an accretion disk as a function of the mass accretion rate. Figure 1 of their
paper shows that one needs $\dot M\ge10^{-5}M_{\sun}/yr$ to get close to the effective
temperature we find in our model. Such high values for $\dot M$ were also found by
\citet{hartmann} and \citet{lachaume2003}. Due to the high accretion rate 
the midplane temperature of the disk in the inner $\approx$10 AU is even much higher than the 
effective disk temperature \citep{bell2}.

Concerning the temperature law farther out than 3 AU the value we adopt here for $q2$ is in good
agreement to what can be found in the literature for isothermal flared disks \citep{kenyon}.  In their model 
the index value diverges from the flat solution where $q=0.75$ at a certain point
in the inner disk and approaches $q=0.5$
for radii much larger than the stellar radius. At intermediate radii the temperature 
index might be approximated by values in between these most extreme cases 
depending on the actual flaring of the disk.

In summary, the temperature distribution indicates that in the innermost disk regions the
disk luminosity is accretion dominated whereas further out it is irradiation dominated. 
This is expected for an active accretion disk since the luminosity released by
accretion is a much steeper function of distance to the star than that by absorption 
of radiation from the central star and the hot innermost disk regions ($L_{accretion}\propto r^{-4}$, whereas
$L_{absorption}\propto r^{-2}$, roughly).
\subsection{Model visibilities}

\subsubsection{MIR results}
For this simple disk model it is straightforward to compute 
the visibilities as a function of projected baseline and wavelength. 
Assuming that each part of the disk surface is emitting as a blackbody 
with an effective temperature $T(r)$ the corresponding total flux and visibility for an 
axisymmetric face-on disk 
is given by \citep[see also equations (2), (3), and following in][]{malbet}:
\begin{equation}
F_{\lambda}=\frac{2\pi}{d^2}\int_{r_{min}}^{r_{max}}rB_{\lambda}(T(r))dr
\end{equation}

\begin{equation}
V_{\lambda}(B_P)=\frac{2\pi}{F_{\lambda}d^2}\int_{r_{min}}^{r_{max}}rB_{\lambda}(T(r))J_0\Bigg[
\frac{2\pi}{\lambda}B_P\frac{r}{d}\Bigg]dr
\end{equation}

with $F_{\lambda}$ being the flux, $B_{\lambda}$ the Planck function, $J_0$ the zeroth-order Bessel function of the first kind, $B_P$ 
the projected baseline and $d$ the distance to the object.

\begin{figure}
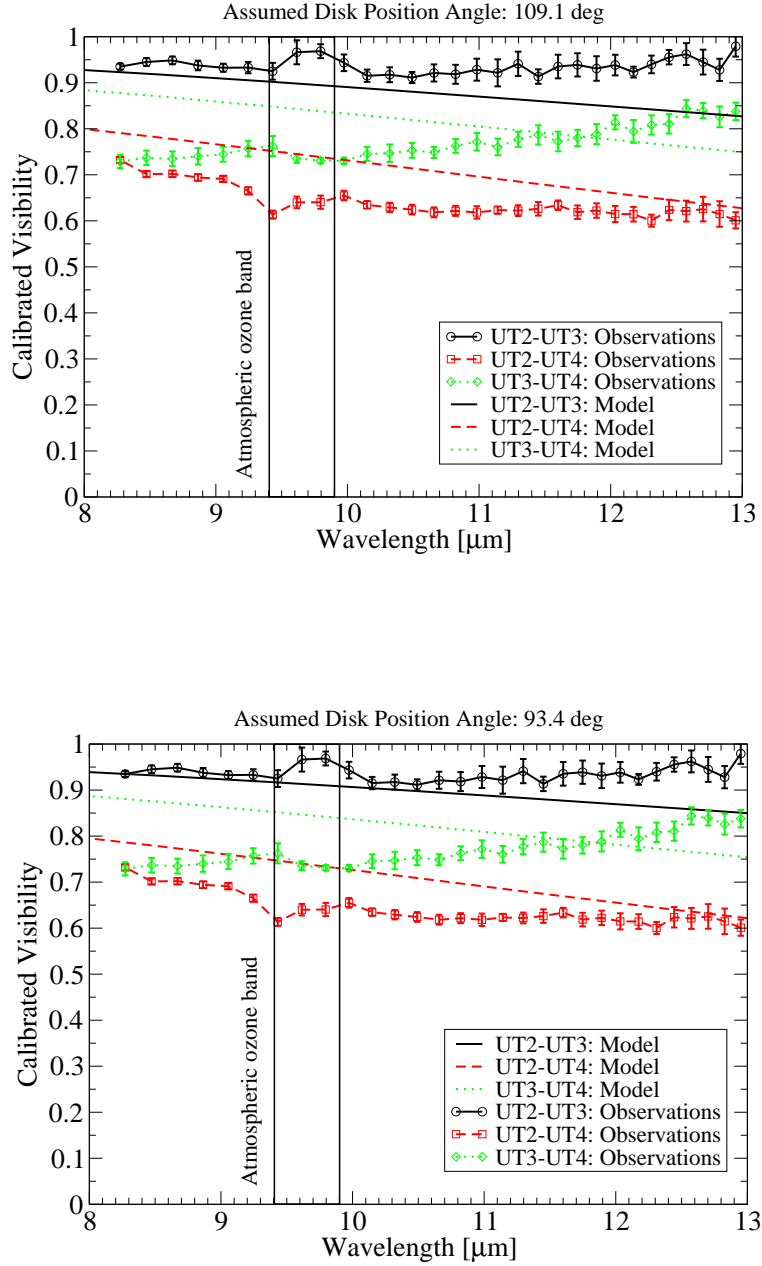

\centering
    \epsscale{0.6}
    \plotone{f10a_color.eps}
     \vspace{2cm}
     \epsscale{0.6}
    \plotone{f10b_color.eps}
   \caption{MIR visibilities derived from the simple disk model in comparison to the observations. 
   The baselines and projection angles
   correspond to those given in Table~\ref{journal}. The disk position angle
   was assumed to be 109.1$^{\circ}$ in the upper panel and 93.4$^{\circ}$ in the lower panel.
   Changing the position angle has the biggest impact on the visibilities of the
   shortest baseline. Apparently, both models do show the
   right trend but they can not account for increasing
   visibilities.}
 \label{model_visi}%
\end{figure}

Taking into account the inclination of the disk and the position angle as given in 
Table~\ref{model_parameter} we computed the corresponding visibilities for the three 
different baselines. The results for two different position angles
are shown in Figure~\ref{model_visi}. In the upper panel the disk inclination and the position
angle are derived from the ellipses fitting in section 5.3. In the 
lower panel the position angle was fitted to the observed visibilities with a least squares fit.
While the ellipse fitting yielded a position angle of 
109.1$^{\circ}\pm$11.6$^{\circ}$ a
least square analysis leads to a value of 93.4$^{\circ}\pm$6.8$^{\circ}$. 
From the plots in Figure~\ref{model_visi} one can see that changing the
position angle mostly affects the visibility of the shortest baseline while 
the curves for the other baselines change only marginally. 

Fitting the UT3-UT4 baseline proofs quite a challenge as the observed visibility is
increasing over the 8-13$\mu$m wavelength interval while all other modeled 
visibilities are decreasing. Especially for this baseline better fits are
expected from more sophisticated disk models.
However, given the simplicity of the current models
the fits seems in reasonable agreement with the observations. 
It clearly shows that interferometric measurements can put additional constraints on 
the structure and geometry of circumstellar disks that cannot be derived from SED 
fitting alone.

\begin{figure}
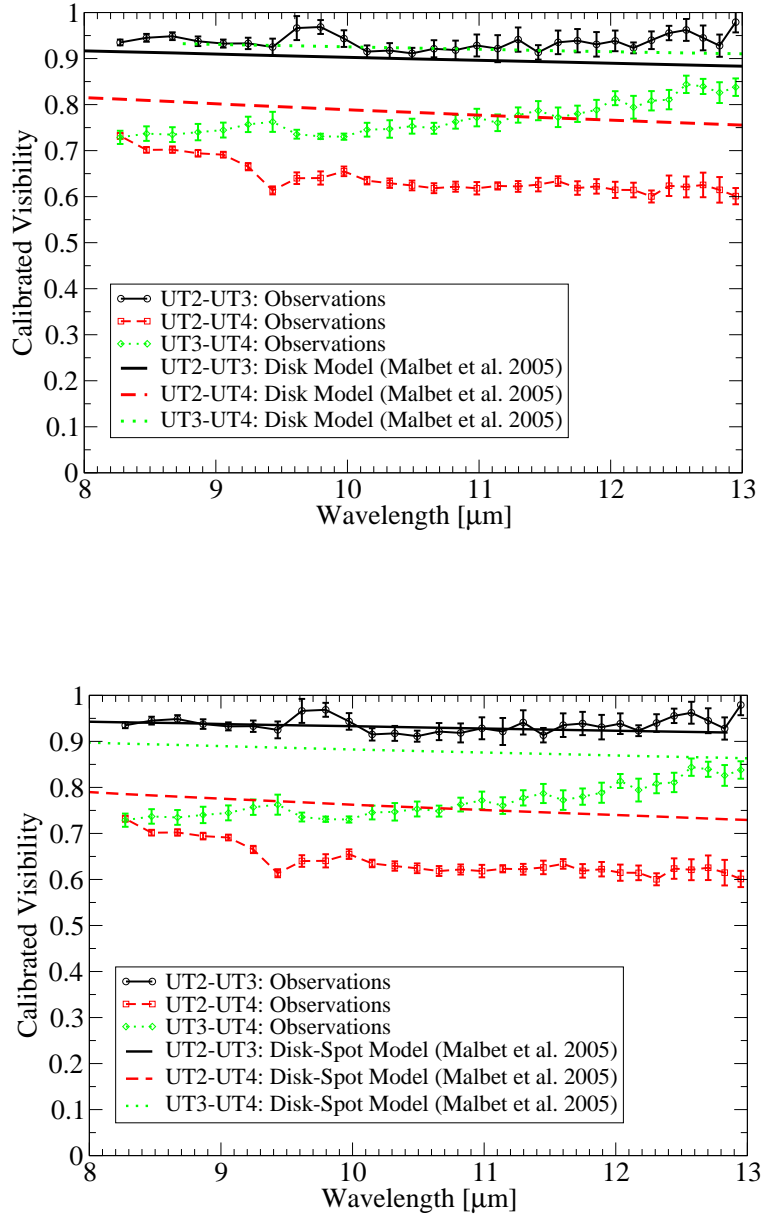

\centering
    \epsscale{0.6}
    \plotone{f11a_color.eps}
     \vspace{2cm}
     \epsscale{0.6}
    \plotone{f11b_color.eps}
   \caption{MIR visibilities derived from models for FU Ori presented in \citet{malbet} in comparison to 
   the observations. \emph{Upper Panel:} Model consisting of a simple accretion disk. This model can be 
   ruled out by our observations as 
   it predicts higher visibilities for the UT3-UT4 baseline than for the UT2-UT3 baseline which is not observed.
   \emph{Lower Panel:}
   Model consisting of an accretion disk and an embedded "hot spot". This model shows the right trend but does not
   fit the
   observations as good as the models presented in this paper.}
 \label{malbetmodels}%
\end{figure}

Interestingly, \citet{malbet} were able to fit their NIR visibilities of FU Ori with a similar 
disk model and also with a disk and an embedded "hot spot". They concluded that the latter
model was more likely based on statistical arguments. The major differences between their 
models and ours is that (1) they fit the 
SED with single power laws for the temperature 
neglecting the flux longwards of 20$\mu$m and (2) their best 
fits yield disk position angles of 47$^{\circ}\,^{+7^\circ}_{-11^\circ}$ and 8$^\circ\pm 21^\circ$. 
The inclination angles (55$^{\circ}\,^{+5^\circ}_{-7^\circ}$ and 48$^{\circ}\,^{+9^\circ}_{-10^\circ}$, respectively), 
are in good agreement with our value.  
In Figure~\ref{malbetmodels} we compare 
both of their models to our observations. We find that both models 
fit the MIDI visibilities not as good as the model presented here. 
Even more we find that our MIR interferometric measurements 
rule out their first model of a simple accretion disk. This model predicts a higher visibility  
for the UT3-UT4 baseline than for the UT2-UT3 baseline which is not observed. 
The model
with an embedded "hot spot" in the accretion disk shows the right trend for the visibilities, but
the values for the UT3-UT4 and UT2-UT4 baseline are higher than for our models which already
give slightly too high values compared to the observations. The main reason for the
higher visibilities resulting from the model applied by \citet{malbet} is the single power law approach
which provides less MIR flux compared to our models.

\subsubsection{NIR results}

We also compared our models to the measured NIR visibilities from \citet{malbet}.
We restricted ourselves to those observations where the object was 
clearly resolved. The results are given in Table~\ref{table_NIRvisi}. 
Except for the observations done with north-west baseline of the Palomar Testbed Interferometer
(PTI/NW) both of our
models agree with the NIR visibilities within the error bars.

 Concerning the predicted "hot spot" in the accretion disk \citep{malbet} our observations do not
 yield new insights into its nature as they do not have the 
 required spatial resolution to confirm its existence. Also the calibrated
 phase of our observations do not contain any information on which basis we could speculate 
 about this possible second companion.

\section{Conclusions and Future Prospects}
We presented the first multi-baseline MIR interferometric observations of
FU Orionis with MIDI/VLTI. The findings can be summarized as follows:
\begin{enumerate}
\item FU Orionis was clearly resolved in the MIR with two VLTI baselines and marginally resolved
with a third baseline indicating the presence of warm dusty material 
surrounding FU Ori out to several tens of AU.

\item The inclination and the position angle of the accretion disk can be inferred
from multi-baseline measurements with MIDI/VLTI. Thus, the instrument provides means
for deriving disk parameters that are otherwise only poorly constrained from 
models and SED fits.
\item The correlated flux indicates that 95\% of the 8-13$\mu$m flux comes
from within the inner 25AU of the disk for the shortest of
our baselines while 65\% of this flux arise from the inner 13AU for our longest baseline.
\item The shapes and strengths of the total 8-13$\mu$m spectrum 
and the (spatially resolved) correlated spectra indicate that most dust particles within the
accretion disk are amorphous and already significantly larger than typical particles
in the ISM.

\item No spectra, neither the total nor the spatially resolved correlated spectra bear significant 
traces of crystalline silicates. Given the high accretion rate and disk temperature
of the system this is unexpected and requires further investigations.

\item The SED and the observed MIR visibilities can be fitted 
reasonably well 
with a simple analytical disk model prescribing a broken power 
law for the effective temperature distribution
of the protoplanetary disk. Within the innermost 3AU the temperature decreases with $T\propto r^{-0.75}$,
farther out the temperature goes with $T\propto r^{-0.53}$. 
The derived power law index for the inner regions
is equal to what is derived theoretically for flat, steady state, optically thick accretion disks \citep{pringle}.
For the outer disk region the derived value is in good agreement to what can be found for isothermal 
flared disks \citep{kenyon}. 

\item Based on points 4. and 5. and recent spectroscopic observations
of protoplanetary disks around TTauri stars and HAeBes it seems questionable 
that FU Ori is a very young classical TTauri star as the size of amorphous dust grains in combination
with the non-detection of crystalline dust contradicts this assumption.  

\item From the acquisition image N-band aperture photometry could be derived for the 
companion FU Ori S. It shows that this young K-type star has more relative MIR 
excess than FU Ori itself hinting at a circumstellar disk with larger flaring angle
or smaller disk inclination.

\end{enumerate}

Putting these results into context with other high spatial 
resolution studies of FU Ori and other FUORs we see that:
\begin{enumerate}
\item Most of the published NIR visibilities of FU Ori \citep{malbet} 
do agree with our newly derived disk models. 
\item Our interferometric observations rule out one out of two disk models for FU Ori 
presented by \citet{malbet}. Due to 
the lack of sufficient spatial resolution the presence
of the proposed second very close companion can neither be confirmed nor 
disproved with the current set of MIDI data. 
\item In contrast to NIR interferometric observations of other FUORs 
\citep{millan-gabet} our
data do not seem to require the introduction of an extended dusty structure providing additional 
uncorrelated flux.
\item For the inner 3AU we require a significantly 
different disk model (e.g., $\Sigma\propto r^{-0.9}$, $T\propto r^{-0.75}$) to fit our visibilities and
SED in comparison to what was found for the outbursting star V1647 
Ori ($\Sigma\propto r^{-1.5}$, $T\propto r^{-0.53}$) \citep{abraham}.
\end{enumerate}

Although the present results provide important new insights into the
FU Ori system new questions arose as well. Future investigations will include
a more thorough and detailed modeling of the observations with 
numerical radiative transfer disk models. These models will not only need to consider the
high accretion rates but also such effects as self-irradiation of the disk.
It will be interesting to see whether these models are able to confirm the 
apparent rotation of the disk's position angle with wavelength and 
whether they can reproduce the increasing visibility curve for the
UT3-UT4 baseline. Also, these disk models will have to reproduce 
the observed NIR and MIR visibilities simultaneously with higher accuracy than
the current simple models are able to do. 
From the observational side re-measuring the UT3-UT4 baseline 
might be eligible and any other new baseline configuration will
certainly help to constrain the disk geometry (especially the position angle) even better.
Concerning the dust structure the 
lack of crystalline silicates needs to be
analyzed in greater detail. Is it still just a contrast effect that we do not see
any crystalline features in the spectra or are they simply not there?
Eventually we will have to face the problem that currently all
FUORs that have been observed with high spatial resolution techniques 
draw a rather inhomogeneous picture of the group. Apparently, some
are still surrounded by nearby dense envelopes while for others
the circumstellar disks suffice to explain the observations. However, even 
those disk-systems (i.e., FU Ori and V1647 Ori) seem to be different in terms of
physical properties as different temperature and surface density profiles are derived.

Without any doubt interferometric observations of circumstellar 
disks in the NIR and MIR provide an unprecedented means for deriving 
geometrical disk 
properties and put new constraints on the physical (and also chemical) processes
taking place within protoplanetary disks. And as more and more interferometric data 
of circumstellar disks are published our view on the cradle of planetary systems 
will be once more refined.



\acknowledgments
S. P. Quanz kindly acknowledges financial support by the German
\emph{Friedrich-Ebert-Stiftung}. Th. Henning and J. Bouwman acknowledge financial support by 
the EC-RTN on "The Formation and 
Evolution of Young Stellar Clusters" ( HPRN-CT-2000-00155).
We are most grateful to R. K\"ohler and F. Lahuis for their support during the data reduction and to 
R. Lachaume and F. Malbet for providing MIR data derived from 
their disk models (section 7.2.). 
We thank our anonymous referee for a detailed and thorough review that 
improved the manuscript. This research has made use of the SIMBAD database,
operated at CDS, Strasbourg, France, and was partly based on observations made with 
the \emph{Spitzer Space Telescope}, which is operated by the Jet Propulsion Laboratory, California Institute of
Technology, under NASA contract 1407.

\clearpage






\begin{deluxetable}{lllll}
\tablecaption{Journal of MIDI observations of FU Orionis and 
the calibrators used for the data reduction. \label{journal}}           
\tablewidth{0pt}
\tablehead{
\colhead{Date} & \colhead{Object} & \colhead{Projected Baseline} & 
\colhead{Position Angle} & \colhead{Comment}
}
\startdata
   31.10.2004 & FU Ori & 44.56 m (UT2-UT3) & 46.54$^\circ$ &  - \\  
              & HD37160&         &               & Calibrator\tablenotemark{a} \\

   02.11.2004 & FU Ori & 86.25 m (UT2-UT4)& 84.24$^\circ$ & - \\
              & HD 31421 &       &               & Calibrator \\
	      & HD 37160 &	 &               & Calibrator \\
	      & HD 50778 &       &               & Calibrator \\

   04.11.2004 & FU Ori & 56.74 m (UT3-UT4)& 106.64$^\circ$& - \\
              & HD 25604 & 	 &               & Calibrator \\
              & HD 20644 & 	 &               & Calibrator \\
              & HD 37160 & 	 &               & Calibrator \\
              & HD 50778 & 	 &               & Calibrator \\

   29.12.2004 & FU Ori & 44.80 m (UT2-UT3)& 46.61$^\circ$ &  -  \\
   	      & HD 37160 &	 &		 & Calibrator \\
	      & HD 94510 &       &               & Calibrator\tablenotemark{a}\\ 
\enddata
\tablenotetext{a}{Observed twice that night.}
\end{deluxetable}

\clearpage

\begin{deluxetable}{llr}
\tablecaption{Photometric values for the FU Ori system. J to L'
are taken from \citet{reipurth}, N-band
values are derived from MIDI 8.7$\mu$m acquisition images. The
errors for the N-band fluxes are standard deviations based on measurements
of three acquisition images. Two were taken on 31.10.2004 and one on 02.11.2004. 
\label{fluxes}}           
\tablewidth{0pt}
\tablehead{
\colhead{} & \colhead{FU Ori} & \colhead{FU Ori S} 
}
\startdata
J [mag] & 6.30$\pm 0.03\quad$ & 10.75$\pm 0.23\quad$  \\
H [mag] & 5.64$\pm 0.05\quad$ & 9.92$\pm 0.21\quad$  \\
K' [mag] & 5.25$\pm 0.02\quad$ & 9.15$\pm 0.15\quad$  \\
L' [mag] & 4.18$\pm0.04\quad$ & 8.09$\pm 0.16\quad$ \\
N [mag] & 2.75$\pm 0.19\quad$ & 5.28$\pm 0.11\quad$\\
\enddata
\end{deluxetable}
\clearpage

\begin{deluxetable}{llll}
\tablecaption{Dust species used in the dust model. Apart from the name,
the chemical formula, the shape and also the reference to laboratory
measurements for the optical properties are given. Mie-theory was used to calculated the
opacities for the homogeneous spheres whereas for the inhomogeneous spheres
we used the distribution of hollow spheres given by \citet{min} to simulate 
grains that are not perfectly symmetric. \emph{References:} (1) \citet{dorschner}, 
(2) \citet{servoin}, (3) \citet{jaeger}, (4) \citet{spitzer}. \label{dust_species}}           
\tablewidth{0pt}
\tablehead{
\colhead{Species} & \colhead{Chemical} & \colhead{Shape} & \colhead{Ref}\\
 \colhead{} & \colhead{Formula} & \colhead{} & \colhead{}
}
\startdata
Amorphous Olivine    & MgFeSiO$_4$         & Homogeneous & (1)  \\
Amorphous Pyroxene & MgFeSi$_2$O$_6$ & Homogeneous & (1)  \\
Crystalline Forsterite  & Mg$_2$SiO$_4$     & Inhomogeneous & (2)  \\
Crystalline Enstatite   & MgSiO$_3$            & Inhomogeneous &  (3)  \\
Amorphous Silica       & SiO$_2$               & Inhomogeneous  &  (4)  \\
\enddata
\end{deluxetable}
\clearpage

\begin{deluxetable}{lllll}
\tablecaption{Possible dust composition for the observed 10$\mu$m silicate 
feature as derived with our dust model. 
The mass fraction of three grain sizes for different dust species is given. \label{dust_compo}}           
\tablewidth{0pt}
\tablehead{
\colhead{} & \colhead{0.1~$\mu$m} & \colhead{1.5~$\mu$m} & \colhead{6.0~$\mu$m} & \colhead{Total}
}
\startdata
Amorphous Olivine & 0.17 & $<$~0.01 & 0.18 & 0.35 \\
Amorphous Pyroxene & $<$~0.01 & 0.07 & 0.57 & 0.64  \\
Crystalline Silicates\tablenotemark{a}  & $<$~0.01 & $<$~0.01 &$<$~0.01 & $<$~0.01 \\
\enddata
\tablenotetext{a}{Forsterite and Enstatite}
\end{deluxetable}
\clearpage

\begin{deluxetable}{cccc}
\tablecaption{Calibrated visibilities for FU Ori observed with three different baselines. 
The errors are standard deviations resulting from calibrations with different 
calibrator stars for each night. \label{visibilities_table}}           
\tablewidth{0pt}
\tablehead{
\colhead{Wavelength [$\mu$m]}  & \colhead{$V$ [UT2-UT3\tablenotemark{a}]} & 
\colhead{$V$ [UT2-UT4\tablenotemark{b}]} & \colhead{$V$ [UT3-UT4\tablenotemark{c}]}
}
\startdata
12.95 & 0.98$\pm$0.02 &0.60$\pm$0.02&0.84$\pm$0.02\\
12.83 &0.93$\pm$0.02 &0.61$\pm$0.03&0.83$\pm$0.02\\
12.70 &0.94$\pm$0.03 &0.63$\pm$0.03&0.84$\pm$0.02\\
12.57 &0.96$\pm$0.02 &0.62$\pm$0.02&0.84$\pm$0.02\\
12.44 &0.96$\pm$0.02 &0.62$\pm$0.02&0.81$\pm$0.02\\
12.31 &0.94$\pm$0.02 &0.60$\pm$0.01&0.81$\pm$0.02\\
12.17 & 0.92$\pm$0.01 &0.61$\pm$0.02&0.79$\pm$0.02\\
12.03 &0.94$\pm$0.02 &0.61$\pm$0.02&0.81$\pm$0.02\\
11.89 &0.93$\pm$0.03 &0.62$\pm$0.02&0.79$\pm$0.02\\
11.75 &0.94$\pm$0.03 &0.62$\pm$0.01&0.78$\pm$0.02\\
11.60 & 0.94$\pm$0.03 &0.63$\pm$0.01&0.77$\pm$0.02\\
11.45 & 0.91$\pm$0.02 &0.63$\pm$0.01&0.79$\pm$0.02\\
11.30 &0.94$\pm$0.03 &0.62$\pm$0.01&0.78$\pm$0.02\\
11.14 &0.92$\pm$0.03 &0.62$\pm$0.01&0.76$\pm$0.02\\
10.98 &0.93$\pm$0.02 &0.62$\pm$0.01&0.77$\pm$0.02\\
10.82 &0.92$\pm$0.02 &0.62$\pm$0.01&0.76$\pm$0.01\\
10.66 &0.92$\pm$0.02 &0.62$\pm$0.01&0.75$\pm$0.01\\
10.49 &0.91$\pm$0.01 & 0.62$\pm$0.01&0.75$\pm$0.02\\
10.32 &0.92$\pm$0.02 &0.63$\pm$0.01&0.75$\pm$0.02\\
10.15 &0.92$\pm$0.01 &0.63$\pm$0.01&0.75$\pm$0.02\\
9.97 &0.94$\pm$0.02 &0.65$\pm$0.01&0.73$\pm$0.01\\
9.80 &0.97$\pm$0.02 &0.64$\pm$0.01&0.73$\pm$0.01\\
9.62 &0.97$\pm$0.03 &0.64$\pm$0.01&0.74$\pm$0.01\\
9.43 & 0.93$\pm$0.02 &0.61$\pm$0.01&0.76$\pm$0.02\\
9.25 & 0.93$\pm$0.01 &0.67$\pm$0.01&0.76$\pm$0.02\\
9.06 & 0.93$\pm$0.01 &0.69$\pm$0.01&0.74$\pm$0.02\\
8.86 &0.94$\pm$0.01 &0.69$\pm$0.01&0.74$\pm$0.02\\
8.67 &0.95$\pm$0.01 &0.70$\pm$0.01&0.73$\pm$0.02\\
8.47 &0.95$\pm$0.01 &0.70$\pm$0.01&0.74$\pm$0.02\\
8.27 &0.94$\pm$0.01 &0.73$\pm$0.01&0.73$\pm$0.01\\
\enddata
\tablenotetext{a}{Projected Baseline: 44.56 m, Position Angle: 46.54$^{\circ}$}
\tablenotetext{b}{Projected Baseline: 86.25 m, Position Angle: 84.24$^{\circ}$}
\tablenotetext{c}{Projected Baseline: 56.74 m, Position Angle: 106.64$^{\circ}$}

\end{deluxetable}
\clearpage

\begin{deluxetable}{llll}
\tablecaption{\emph{Upper half:} Derived FWHM (in AU) of Gaussian brightness distributions for
the three baselines and three wavelengths using equation~\ref{visi_eq}. The distance to FU Ori is 
assumed to be 450 pc. \emph{Lower half:} Parameters of best ellipses fitted to the data for the  
three considered wavelengths.\label{table_sizes}}           
\tablewidth{0pt}
\tablehead{
\colhead{} & \colhead{9.0~$\mu$m} & \colhead{11.0~$\mu$m} & \colhead{12.5~$\mu$m} 
}
\startdata
UT2-UT3 (44.56 m) & 2.47$^{+0.40}_{-0.46}$ & 3.27$^{+0.67}_{-0.82}$ & 3.43$^{+0.81}_{-1.02}$ \\
UT2-UT4 (86.25 m) & 3.25$^{+0.24}_{-0.25}$ & 4.34$^{+0.07}_{-0.08}$ & 4.93$^{+0.17}_{-0.17}$  \\
UT3-UT4 (56.74 m) & 4.19$^{+0.28}_{-0.30}$ & 5.00$^{+0.35}_{-0.37}$ & 4.83$^{+0.43}_{-0.46}$ \\
\hline
Semimajor axis $a$ [AU]& 2.28  &  2.51 & 2.54\\ 
Semiminor axis $b$ [AU]& 1.21  &  1.54 & 1.42\\
Ellipse area [AU$^2$]& 8.67  &  12.14 & 11.33\\
Position angle $\theta$\tablenotemark{a} & 122.2$^{\circ}$ & 111.1$^{\circ}$ & 93.9$^{\circ}$\\
Inclination $i$\tablenotemark{b}& 58.0$^{\circ}$ & 52.2$^{\circ}$ & 56.0$^{\circ}$\\
\enddata
\tablenotetext{a}{Position of $a$ measured from north eastwards}
\tablenotetext{b}{Assuming an underlying circular disk}
\end{deluxetable}

\clearpage

\begin{deluxetable}{lll}
\tablecaption{Parameters of the simple disk model used to fit the SED. 
The upper part of the table gives the parameters for the inner 3 AU. Farther out
in the disk the temperature and surface density follow distributions defined
by the parameters in the lower part of the table. \label{model_parameter}}           
\tablewidth{0pt}
\tablehead{
\colhead{Parameter} & \colhead{Variable} & \colhead{Value}  
}
\startdata
Inner disk radius& $R_{in}$ & $5.5~R_{\sun}$ \\
Outer disk radius& $R_{out}$ & 100 AU\\
Disk inclination~\tablenotemark{a}& $i$ & $55.4^{\circ}$\\
Disk position angle~\tablenotemark{b}& $\theta$ & $109.1^{\circ}$~\tablenotemark{a}~/~$93.4^{\circ}$~\tablenotemark{c}\\
Extinction& A$_V$ & $2.6$\\
Temperature at 1 AU& $T_{1AU,1}$& $670~K$\\
Power law index for temperature& $q1$ &  $-0.75$\\
Power law index for surface density& $p1$ & $-0.9$\\
Disk mass& $M_{D,1}$& $0.03~M_{\sun}$\\ \hline

Temperature at 1 AU& $T_{1AU,2}$& $550~K$\\
Power law index for temperature& $q2$ &  $-0.53$\\
Power law index for surface density& $p2$ & $-1.4$\\
Disk mass& $M_{D,2}$& $0.01~M_{\sun}$\\
\enddata
\tablenotetext{a}{As derived in section 5.3.}
\tablenotetext{b}{Only required for visibility computation}
\tablenotetext{c}{Resulting from $\chi^2$-fit}
\end{deluxetable}

\clearpage

\begin{deluxetable}{lcllcc}
\tablecaption{Comparison of observed NIR square visibilities from \citet{malbet} to our disk model
with two different position angles. The observations were carried out with different baselines 
at the Palomar Testbed Interferometer (PTI) and the Very Large Telescope Interferometer (VLTI).\label{table_NIRvisi}}           
\tablewidth{0pt}
\tablehead{
\colhead{Interferometer} & \colhead{Filter} & \colhead{Baseline [m]}  &
\colhead{Pos. Angle [deg]} & \colhead{$V^2$ (observed)} & \colhead{$V^2$ (model)\tablenotemark{a}}
}
\startdata
PTI/NS & H & 103.7 & 61.8 & 0.83$\pm$0.04 & 0.80 / 0.82 \\
PTI/NW & H & 85.6  & 14.1 & 0.79$\pm$0.05 & 0.92 / 0.94\\
PTI/NS & K & 102.7 & 63.6 & 0.72$\pm$0.08 & 0.78 / 0.80\\
PTI/NW & K & 84.2  & 14.0 & 0.79$\pm$0.05 & 0.91 / 0.93\\
PTI/SW & K & 82.5  & -55.6& 0.82$\pm$0.08 & 0.85 / 0.87\\
VLTI/UT1-UT3\tablenotemark{b} & K  & 89.6 & 54.2 & 0.87$\pm$0.05 & 0.83 / 0.85 \\
\enddata
\tablenotetext{a}{For two different position angles: 109.1$^{\circ}$~/~93.4$^{\circ}$}
\tablenotetext{b}{VINCI data}

\end{deluxetable}

\clearpage


  
\clearpage


\clearpage

\clearpage

\clearpage

\clearpage

\clearpage

\clearpage

\clearpage

\clearpage



\clearpage

\begin{thebibliography}{}


\bibitem[Adams, Lada \& Shu(1987)]{adams} 
Adams, F.C., Lada, C.J., Shu, F.H.: ApJ \textbf{312}, 788 (1987)

\bibitem[\'Abrah\'am et al.(2006)]{abraham} 
\'Abrah\'am, P., Mosoni, L., Henning, Th., K\'osp\'al, \'A., Leinert, Ch., Quanz, S.P., 
Ratzka, Th.: A\&A \textbf{449}, L13 (2006)

\bibitem[Bell et al.(1995)]{bell} 
Bell, K.R., Lin, D.N.C., Hartmann, L.W., Kenyon, S.J.: ApJ \textbf{444}, 376 (1995)

\bibitem[Bell et al.(1997)]{bell2} 
Bell, K.R., Cassen, P.M., Klahr, H.H., Henning, Th.: ApJ \textbf{486}, 372 (1997)

\bibitem[Bonnell \& Bastien(1992)]{bonnell} 
Bonnell, I. \& Bastien, B.: ApJ \textbf{401}, L31 (1992)

\bibitem[Bouwman et al.(2001)]{bouwman} Bouwman, J., Meeus, G., de Koter, A., Hony, S., Dominik, C.,
Waters, L.B.F.M.: A\&A \textbf{375}, 950 (2001)

\bibitem[Bouwman et al.(2006)]{bouwman2006} 
Bouwman, J., Henning, Th., et al. (2006, in preparation)

\bibitem[Clarke et al.(2005)]{clarke} 
Clarke, C., Lodato, G., Melnikov, S.Y., Ibrahimov, M.A.: MNRAS \textbf{361}, 942 (2005)

\bibitem[Donati et al.(2005)]{donati} 
Donati, J.-F., Paletou, F., Bouvier, J., Ferreira, J.: Nature \textbf{438}, 466 (2005)

\bibitem[Dorschner et al.(1995)]{dorschner} 
Dorschner, J., Begemann, B., Henning, Th., J\"ager, C., and Mutscke, H.: A\&A \textbf{300}, 503 (1995)

\bibitem[Ducati(2002)]{ducati} 
Ducati, J.R.: VizieR On-line Data Catalogue: II/237 (2002)

\bibitem[Dullemond, Apai \& Walch(2006)]{dullemond}
Dullemond, C.P., Apai, D. \& Walch, S.: A\&A \textbf{640}, L78 (2006)

\bibitem[Gail(2004)]{gail1} 
Gail, H.-P.: A\&A \textbf{413}, 571 (2004)

\bibitem[Gail(2001)]{gail2} 
Gail, H.-P.: A\&A \textbf{378}, 192 (2001)

\bibitem[Gullbring et al.(1998)]
{gullbring} Gullbring, E., Hartmann, L., Briceno, C., Calvet, N.: ApJ \textbf{492}, 323 (1998)

\bibitem[Haisch et al.(2001)]{haisch}
Haisch, K.E.Jr., Lada, E.A., Lada, C.J.: AJ \textbf{121}, 1512 (2001)

\bibitem[Hanner et al.(1998)]{hanner} 
Hanner, M.S., Brooke, T.Y., Tokunaga, A.T.: ApJ \textbf{502}, 871 (1998)

\bibitem[Hartmann \& Kenyon(1996)]{hartmann} 
Hartmann, L. \& Kenyon, S.J.: ARAA \textbf{34}, 207 (1996)

\bibitem[Henning et al.(2006)]{henning2006} 
Henning, Th., Mutschke, H. \& J\"ager, C. (2006),
In \emph{Astrochemistry: Recent Successes and Current Challenges},
(D.C. Lis, G.A. Blake \& E. Herbst, eds), Proceedings IAU Symposium 231


\bibitem[Herbig et al.(2003)]
{herbig} Herbig, G.H., Petrov, P.P., Duemmler, R.: ApJ \textbf{595}, 384 (2003)

\bibitem[Herbig(1977)]
{herbig1977} Herbig, G.H.: ApJ \textbf{217}, 693 (1977)

\bibitem[Higdon et al.(2004)]{higdon}
Higdon, S.J.U., Devost, D., Higdon, J.L., Brandl, B.R., Houck, J.R., et al.: PASP \textbf{116}, 975 (2004)

\bibitem[J\"ager et al.(1998)]{jaeger}
J\"ager, C., Molster, F.J., Dorschner, J., et al.: A\&A \textbf{339}, 904 (1998)

\bibitem[Kenyon \& Hartmann(1987)]{kenyon}
Kenyon, S.J., Hartmann, L.:  ApJ \textbf{323}, 714 (1987)

\bibitem[Kenyon, Hartmann \& Hewett(1988)]{kenyon1988}
Kenyon, S.J., Hartmann, L. \& Hewett, R.:  ApJ \textbf{325}, 231 (1988)

\bibitem[Lachaume(2004)]{lachaume} 
Lachaume, R.: A\&A \textbf{422}, 171 (2004)

\bibitem[Lachaume et al.(2003)]{lachaume2003} 
Lachaume, R., Malbet, F., Monin, J.-L.: A\&A \textbf{400}, 185 (2003)

\bibitem[Leinert et al.(2004)]{leinert} 
Leinert, Ch., van Boekel, R. et al.: A\&A \textbf{537}, 423 (2004)

\bibitem[Lodato \& Clarke(2004)]{lodato} 
Lodato, G. \& Clarke, C.J.: MNRAS \textbf{353}, 841 (2004)

\bibitem[Malbet et al.(2005)]{malbet} Malbet, F., Lachaume, R., Berger, J.-P., Colavita, M. M., di Folco, E.,
Eisner, J. A., Lane, B. F., Millan-Gabet, R., Segransan, D., Traub, W. A.: A\&A \textbf{437}, 627 (2005)

\bibitem[Mathis(1990)]{mathis}
Mathis, J.S.: ARA\&A \textbf{28}, 37 (1990)

\bibitem[Mennesson et al.(2005)]{mennesson}
Mennesson, B., Koresko, C., Creech-Eakman, M. J., Serabyn, E., Colavita, M. M., Akeson, R., Appleby, E., Bell, J., Booth, A., Crawford, S., et al.: ApJ \textbf{634}, L169 (2005)

\bibitem[Men'shikov \& Henning(1997)]{henning} 
Men'shchikov, A.B. \& Henning, Th.: A\&A \textbf{879}, 318 (1997)

\bibitem[Millan-Gabet et al.(2006)]{millan-gabet} 
Millan-Gabet, R., Monnier, J.D., Akeson, R.L., Hartmann, L., 
Berger, J.-P., et al.: astro-ph/0512230 (accepted by ApJ)

\bibitem[Min et al.(2005)]{min} 
Min, M., Hovenier, J.W. \& de Koter, A.: A\&A \textbf{432}, 909 (2005)

\bibitem[Molster \& Kemper(2005)]{molster} 
Molster, F. \& Kemper, C.: Space Science Reviews \textbf{119}, 3 (2005)

\bibitem[Peeters et al.(2002)]{peeters} 
Peeters, E., Hony, S., Van Kerckhoven, C., et al.: A\&A \textbf{390}, 1089 (2002)

\bibitem[Pringle (1981)]{pringle} 
Pringle, J.E.: ARAA \textbf{19}, 137 (1981)

\bibitem[Przygodda(2004)]{przygodda} 
Przygodda, F.: PhD Thesis at the University of Heidelberg (2004)


\bibitem[Schegerer et al.(2006)]{schegerer}
Schegerer, A., Wolf, S., et al. (2006, in preparation)

\bibitem[Reipurth \& Aspin(2004)]{reipurth} 
Reipurth, B. \& Aspin, C.: ApJ \textbf{608}, L65 (2004)

\bibitem[Servoin \& Piriou(1973)]{servoin} 
Servoin, J.L. \& Piriou, B.: phys. stat. sol. \textbf{55}, 677 (1973)

\bibitem[Spitzer \& Kleinman(1960)]{spitzer} 
Spitzer, W.G. \& Kleinman, D.A.: Physical Review \textbf{121}, 1324 (1960)

\bibitem[van Boekel et al.(2004)]{boekel} 
van Boekel, R., Min, M., Leinert, Ch., Waters, L.B.F.M., Richichi, A., Chesneau, O., Dominik, C., Jaffe, W., 
et al.: Nature \textbf{432}, 479 (2004)

\bibitem[van den Ancker(2000)]
{ancker} van den Ancker, M.E., Bouwman, J., Wesselius, P.R., et al.: A\&A \textbf{357}, 325 (2000)


\bibitem[van Diedenhoven(2004)]
{diedenhoven} van Diedenhoven, B., Peeters, E., Van Kerckhoven, C., et al.: ApJ \textbf{611}, 928 (2004)

\bibitem[Voshchinnikov et al.(2006)]{voshchinnikov}
Voshchinnikov, N. V., Il'in, V. B., Henning, Th., Dubkova, D. N.: A\&A \textbf{445}, 167 (2006)

\bibitem[Wang et al.(2004)]{wang}
Wang, H., Apai, D., Henning, Th., Pascucci, I.: ApJ \textbf{601}, L83 (2004)

\bibitem[Weintraub et al.(1991)]{weintraub} 
Weintraub, D.A., Sandell, G., Duncan, W.D.: ApJ \textbf{382}, 270 (1991)

\bibitem[Welin(1971)]
{welin} Welin, G.: A\&A \textbf{12}, 312 (1971)


\end{thebibliography}
\end{document}